\documentclass []{aa}
\usepackage{graphicx}
\usepackage{txfonts}
\begin{document}
   \title{Direct evidence of the receding `torus' around central
   nuclei of powerful radio sources} 

   \titlerunning{Direct evidence of the receding 'torus' in powerful
   radio sources}


   \author{T. G. Arshakian\inst{1,2}
   } \institute{Max-Planck-Institut f\"ur Radioastronomie (MPIfR), Auf
   dem H\"ugel 69, 53121 Bonn, Germany \\
   \email{tigar@mpifr-bonn.mpg.de} \and Byurakan Astrophysical
   Observatory, Aragatsotn prov. 378433, Armenia and Isaac Newton
   Institute of Chile, Armenian Branch }

   \date{17 September 2004}

   \abstract{The presence of obscuring material (or a dusty `torus')
   in active galactic nuclei (AGN) is central to the unification model
   for AGN.  Two models, the multi-population model for radio
   sources and the receding torus model, are capable of describing 
   observational properties of powerful radio galaxies and radio
   quasars. Here, I study the changes of the opening angle of the
   obscuring torus with radio luminosity at 151 MHz, [O III]
   emission-line luminosity and cosmic epoch aiming to discriminate
   between two working models. An analytical expression relating the
   half opening angle of the torus to the mean projected linear sizes
   of double radio galaxies and quasars is derived. The sizes of
   powerful double radio sources taken from the combined sample of
   3CRR, 6CE and 7CR complete samples are used to estimate the torus
   opening angle. I found a statistically significant correlation
   between the half opening angle of the torus and [O III]
   emission-line luminosity. The opening angle increases from
   20$^{\circ}$ to 60$^{\circ}$ with increasing [O III] emission-line
   luminosity. This correlation is interpreted as direct evidence of
   the receding torus around central engines of powerful double radio
   sources.

      }

   \maketitle
%

\section{Introduction}
Spectropolarimetric observations of Seyfert 2 galaxies (Antonucci \&
Miller 1985; Miller \& Goodrich 1990) suggested an orientation-based
unification scheme for active galactic nuclei (AGN). An optically
thick molecular torus obscuring the AGN permits a unification of
Seyfert 1 and Seyfert 2 galaxies by proposing that they are
intrinsically the same objects viewed at different angles to the axis
of the torus. This model was invoked by Scheuer (1987) and Barthel
(1989) to account for the orientation-dependent appearance of powerful
radio sources, which can appear as radio galaxies or
quasars. According to the standard picture, the torus hides the AGN
and broad-line region of a radio galaxy, whereas for quasars these are
viewed directly for a quasar down the ionization cone. The opening
angle of the torus defines the cone angle and the half opening angle
is the critical angle ($\theta_{\rm c}\sim45^{\circ}$; Barthel 1989)
at which the transition between radio galaxies and quasars occurs. The
narrow-line regions have extended kpc-scale structure and therefore
are visible for any orientation of the torus. There is considerable
multi-band observational evidence supporting the unification scheme
for radio galaxies and quasars. This comes from the X-ray detection of 
inverse Compton scattered photons (Brunetti et al. 1997) and polarized
(scattered) emission lines from hidden quasars in some radio galaxies
(Antonucci \& Miller 1985), from the direct detection of unabsorbed soft
and hard X-ray emissions from quasars (e.g. 3C 109, Allen and Fabian
1992; Cyg A, Ueno et al. 1994; 3C 295, Brunetti et al. 2001; 3C 265,
Bondi et al. 2004), from the detection of comparable far-infrared
emission from the high-redshift ($z>0.8$) radio galaxies and quasars
(e.g. Meisenheimer et al. 2001; Haas et al. 2004), from the difference
of the mid-infrared spectral energy distribution in the broad-line and
narrow-line AGN (Siebenmorgen et al. 2004), from the detection of
superluminal radio knots on parsec-scales (Vermuelen \& Cohen 1994),
and from the presence of supermassive black holes and relativistic
jets in both radio galaxies and quasars (McLure \& Dunlop 2002;
Laing 1988; Garrington et al. 1988).

A single opening angle can not account for two pieces of observational
evidence: (i) at a given radio power the [O III] lines are weaker in
radio galaxies than in quasars (Lawrence 1991; Simpson 2003), and (ii)
the increase of the quasar fraction with increasing radio luminosity
(Hill et al. 1996; Willott et al. 2000; Grimes et al. 2004). Lawrence
(1991) suggested the receding torus model in which the dust
evaporation radius of the torus depends on the luminosity of the
photoionizing radiation of the AGN, which is assumed to be independent
of the height of the torus. In this scenario the opening angle of the
torus increases with increasing luminosity of the ionizing radiation,
$L_{\rm phot}$, from the AGN. This model accounts for the fact that
quasars are brighter in $L_{\rm phot}$ than radio galaxies (Lawrence
1991; Simpson 2003) in samples selected at low radio frequencies, and
that the fraction of quasars increases with radio luminosity. 
Gopal-Krishna et al. (1996) showed that a wider opening angle with
increasing radio luminosity could explain the problem of unification
posed by the difference between the radio luminosity-size correlations
for radio galaxies and quasars. In addition, Willott et al. (2000)
found that the fraction of quasars decreases at low luminosities. They
argued that this can be explained either by the receding torus model
(decrease of $\theta_{\rm c}$ with decreasing radio luminosity), or by
the existence of a second population of low-luminosity radio
galaxies. Grimes et al. (2004) used a combined complete sample at 151
MHz to show that, in the luminosity-dependent unified scheme, a
two-population model may explain the increase of the quasar fraction
with $L_{\rm phot}$ as well as the emission-line differences between
radio galaxies and quasars. They conclude that the effect of the
receding torus may be important but it is not clear yet whether the
receding torus is present in both populations.

The opening angle of the torus, $\theta_{\rm c}$, defines the geometry
of the torus (the inner radius and height) and appears to be an
important parameter for understanding the evolutionary behaviour of the
obscuring material and testing the torus model. In the Appendix, I
derive an equation that allows the mean half opening angle of the
torus $\overline{\theta}_{\rm c}$ to be estimated by means of the
projected linear sizes of radio galaxies and quasars. In section 2
the equations for estimating $\overline{\theta}_{\rm c}$ are
presented. The combined complete sample of radio sources is described
in section 3. In section 4, the relations between the critical angle with
the 151 MHz radio luminosity $L_{151}$, [O III] 5007-\AA\ emission-line
luminosity $L_{[\rm O\,III] }$ and cosmic epoch are investigated. In
section 5, the receding torus model is discussed.

Throughout the paper a flat cosmology (${\it \Omega}_{\rm m} + {\it\Omega}_{\rm
 \Lambda}= 1$) with non-zero lambda, ${\it \Omega}_{\rm m} = 0.3$ and 
 {\it H}$_{\rm 0} = 70\,\, {\rm km\,\,s}^{-1}\,{\rm Mpc}^{-1}$ is used.

\section{Equations for the opening angle of the torus}
Powerful FRII (Fanaroff \& Riley 1974) radio galaxies and quasars are
associated with bipolar relativistic radio jets emerging from the AGN
which form extended radio lobes on both sides of the AGN. The radio
jets are almost aligned with the axes of the tori in FRII (Cyg A;
Canalizo et al. 2004) and FRI radio sources (3C270, Jaffe et al. 1993;
Verdoes Kleijn et al. 2001). The half opening angle of the torus is
defined by the angle between the radio axis and the direction at which
the division between radio galaxies and quasars occurs. Then the half
opening angle of the torus can be estimated by means of the projected
linear sizes of FRII radio galaxies and quasars; linear sizes of
quasars appear to be systematically smaller than the linear sizes of
radio galaxies (Barthel 1989) as a result of projection effects. In
the Appendix, I derive an equation (7) for estimating the mean
critical angle $\overline{\theta}_{\rm cs}$ from the ratio of the mean
projected linear sizes of quasars $\overline{l}_{\rm Q}$ and the whole
sample $\overline{l}$. The problem is solved for the sample of FRII
radio sources having an isotropic distribution of radio axes over the
sky. Low-frequency radio samples of radio sources are thought to be
free from orientation biases and can be used in this analysis.

Another independent way to estimate the half opening angle is to
consider the number of quasars $N_{\rm Q}$ and the number of radio
galaxies and quasars $N_{\rm {G+Q}}$ in the low-frequency radio
samples,
\begin{equation}
  \overline{\theta}_{\rm cf} = \arccos \left(1 - \frac{N_{\rm
  Q}}{N_{\rm G+Q}} \right).
\end{equation}
If the unified schemes for FRII radio galaxies and quasars is valid
then one should expect a correlation between $\overline{\theta}_{\rm
cf}$ and $\overline{\theta}_{\rm cs}$, which are independent estimates.

   \begin{table}
     \caption[]{The Kolmogorov-Smirnov test is used to estimate
the significance level of the hypothesis that the [O\,III]
emission-line luminosity (and radio luminosity) distributions of HEGs
and quasars (QSOs), HEGs and WQs, and, QSOs and WQs are drawn from the
same parent distribution.}
     \label{KS-test}
     $$ 
     \begin{tabular}{cccc}
       \hline
       \noalign{\smallskip}
       $L_{[\rm {O III}]}$ & {K-S\,test} & $L_{151}$ & {K-S\,test}  \\
       distribution        &   (\%)      & distribution   &   (\%)       \\
       \noalign{\smallskip}
       \hline
       \noalign{\smallskip}
       QSO-HEG & 99.99 & QSO-HEG & 99.66  \\
       QSO-WQ  & 98.77 & QSO-WQ  & 99.81  \\
       HEG-WQ  & 16.60 & HEG-WQ  & 79.47  \\
       \noalign{\smallskip}
       \hline
     \end{tabular}
     $$ 
   \end{table}

\section{The sample and selection criteria}
I use a combined sample of three complete low-frequency samples
(Grimes et al. 2004)\footnote{The electronic version of the combined
sample is available at
http://www-astro.physics.ox.ac.uk/$\sim$sr/grimes.html.}, made up of 
the 3CRR, 6CE and 7CRS samples (Laing, Riley \& Longair 1983;
Rawlings, Eales \& Lacy 2001; Willott et al. 2002). This contains
302 radio sources all having extended radio structure on
kiloparsec-scales. I then excluded 3C48, 3C287 and 3C343 because
they are complex with no clear double lobe structure (Laing et
al. 1983). Three other sources were excluded: 3C231 because its radio
emission is due to a starburst, 3C345 and 3C454.3 because their
inclusion in the 3CRR sample is a result of Doppler boosting of the
jet.

I adopted the classification of radio sources used by Grimes et
al. (2004): high- and low-excitation narrow-line radio galaxies (HEGs
and LEGs), weak quasars (WQs) and broad-line radio quasars.  The
category of weak quasars includes HEGs which are objects with weakly
or heavily obscured broad-line nuclei seen indirectly (e.g., in broad
wings of H${\alpha}$ and Paschen lines, in optically-polarized light)
and unobscured objects with weak broad-line optical nuclei.

There are 39 low-excitation radio galaxies in the sample which
are believed to be a separate population independent from the
high-excitation radio galaxies and radio quasars (Laing et
al. 1994). Exclusion of these sources and FRI radio sources leaves a
sample consisting of 237 FRII radio sources. The projected linear
sizes (the apparent distance between radio lobes) of 3CRR
sources are adopted from the 3CRR database\footnote{The electronic
version of the 3CRR database can be found at
http://www.3crr.dyndns.org/cgi/database}, and the sizes of 6CE and
7CRS sources are taken from the electronic version of the combined
database$^1$. The projected linear sizes vary over (1 to 2000) kpc
over the range $(1.6\times10^{24}$ to $4.6\times10^{28})$ W Hz$^{-1}$
sr$^{-1}$ of the 151 MHz radio luminosity.

   \begin{figure}
   \centering
   \includegraphics[angle=-90,width=8.2cm]{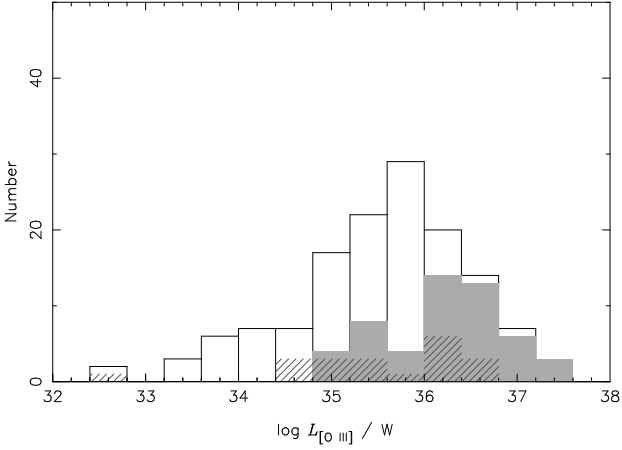}
      \caption{The distribution of [O\,III] emission-line
      luminosity for 134 high-excitation narrow-line galaxies (open
      area), 52 quasars (grey area) and 20 weak quasars (hatched area)
      in the combined 3CRR, 6CE and 7CRS samples.}
         \label{OIII-hist}
   \end{figure}
%
   \begin{figure}
   \centering
   \includegraphics[angle=-90,width=8.2cm]{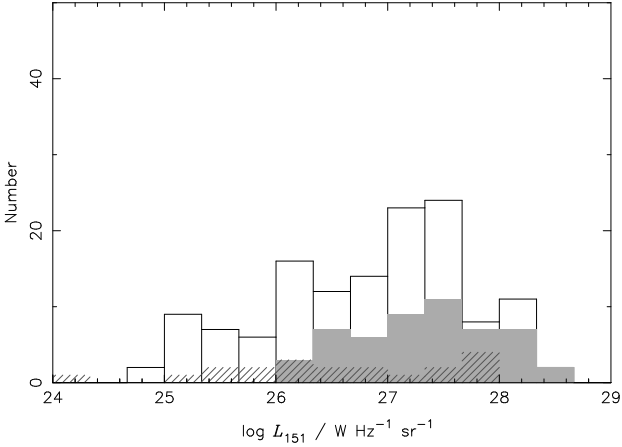}
      \caption{The distribution of radio luminosity at 151 MHz for 134
      HEGs (open area), 52 quasars (grey area) and 20 WQs (hatched
      area) in the combined sample. }
         \label{151-hist}
   \end{figure}
%

   \begin{figure}
   \centering
   \includegraphics[angle=-90,width=8.2cm]{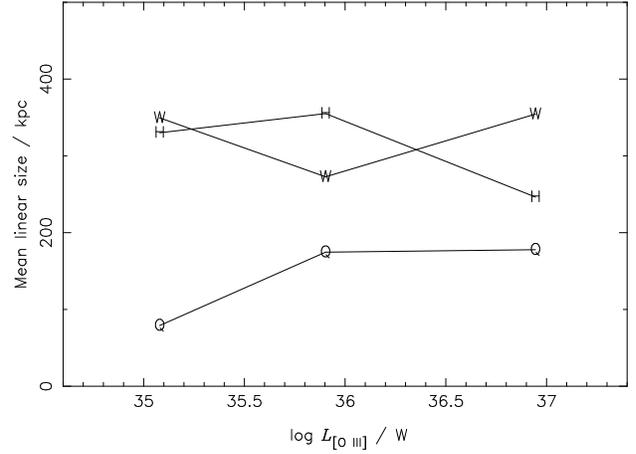}
      \caption{Mean projected linear sizes of 206 FRII radio sources
      are calculated for three equal subsamples binned by [O III]
      emission-line luminosity. Quasars, weak quasars and HEGs are
      denoted by 'Q', 'W' and 'H' respectively. }
         \label{size_LOIII}
   \end{figure}
%

The structure of FRII radio sources is influenced by environmental
asymmetry, which becomes more important on small scales (Arshakian \&
Longair 2000). I excluded the compact steep-spectrum radio sources
with typical sizes $<$ 20 kpc to avoid contamination of projection
effects by interaction effects between a radio jet and interstellar
medium (Barthel 1989). Their inclusion or exclusion does not introduce
perceptable change in the results described below.  The final sample
consists of 206 FRII radio sources: 134 HEGs, 20 weak quasars and 52
quasars.

\begin{table*}
  \caption[]{Estimates of critical angles, ${\theta}_{\rm cs}$ and
  ${\theta}_{\rm cf}$ and their errors in equal subsamples binned by
  $L_{[\rm {O III}]}$. The first line of the table refers to the two
  sets of binned data, and the second line shows results of three sets
  of binned data. }
  \label{F-test}
  $$ 
  \begin{array}{ccccccccccc}
    \hline
    \noalign{\smallskip}
    (1) &  (2) & (3) & (4) & (5) & (6) & (7) & (8) & (9) & (10) & (11) \\
    \hline
    \noalign{\smallskip}
    N_{\mathrm {bin}}  &  \log (L_{[\mathrm {O III}]}/{\rm W}) &  \log ({\overline L}_{[\mathrm {O III}]}/{\rm W}) & N_{\mathrm {G+Q}} & \,\,\,N_{\mathrm Q}&  {\theta}_{\mathrm {cf}} \pm \sigma_{{\theta}_{\mathrm {cf}}} & Fisher \,\mathrm{test} & \overline{l} \pm \sigma_{\overline {l}}  & \overline {l}_{\rm Q} \pm \sigma_{\overline {l}_{\rm Q}} \,\,\, & \,\,\,{\theta}_{\mathrm {cs}} \pm \sigma_{{\theta}_{\mathrm {cs}}} & Fisher \,\mathrm{test} \\

     & \mathrm {bin \,\,width} & & & & ({^\circ}) &   (\%)& (\mathrm {kpc}) & (\mathrm {kpc})  & ({^\circ}) & (\%) \\
    \noalign{\smallskip}
    \hline
    2 & 32.43-35.81 & 35.51 & 103 & 12 & 28 \pm 35 & 48.6 & 341 \pm 36 &  111 \pm 25 & 22 \pm 6  & 99.4 \\
      & 35.81-37.37 & 37.08 & 103 & 40 & 52 \pm 12 &      & 217 \pm 19 &  177 \pm 22 & 63 \pm 14 &      \\ 
    \hline
    3 &  32.43-35.38 & 35.08 & 68 & 6  & 24 \pm 57 & 17.5 & 311 \pm 11 &  79 \pm 19  & 17 \pm 5 & 97.4  \\
      &  35.38-36.13 & 35.90 & 68 & 15 & 39 \pm 24 &      & 308 \pm 43 & 175 \pm 39 & 41 \pm 12 &       \\
      &  36.13-37.37 & 37.09 & 70 & 31 & 56 \pm 13 &      & 225 \pm 23 & 178 \pm 25 & 59 \pm 14 &       \\
    \noalign{\smallskip}
    \hline
  \end{array}
  $$ 
  \begin{list}{}{}
  \item Columns: (1) - The number of bins; (2) - [O III] luminosity
  bin width; (3) - Logarithm of the central value of the $L_{[\mathrm
  {O III}]}$ in the bin; (4,5) - Number of radio galaxies and quasars
  in the $\log L_{[\mathrm {O III}]}$ bin; (6) - The half opening
  angle of the torus estimated by fraction of quasars (eq. 1), and its
  standard error $\sigma_{{\theta}_{\mathrm {cf}}}$ is calculated
  assuming that $\sigma_{N_{\rm G+Q}} = \sqrt{N_{\rm G+Q}}$ and
  $\sigma_{N_{\rm Q}}=\sqrt{N_{\rm Q}}$; (7) - The confidence level of
  rejecting the null hypothesis that half opening angles
  ${\theta}_{\mathrm {cf}}$ of the torus estimated in different
  $L_{[\rm O\,III]}$ bins are equal. The variance analysis (the
  $Fisher$ test) is used with two degrees of freedom, $N_{\rm bin}-1$
  and $N_{\rm bin}(N_{\rm G+Q}-1)$; (8,9) - The mean linear sizes of
  all sources (radio galaxies and quasars) and quasars, and their
  standard errors; (10) - The half opening angle of the torus and its
  standard error from eqs, (7) and (8); (11) - The confidence level
  for rejecting the null hypothesis that half opening angles
  ${\theta}_{\mathrm {cs}}$ of the torus are equal in different
  $L_{[\rm O\,III]}$ bins.
  \end{list}
\end{table*}

\section{Relations for the torus opening angle}
The distribution of [O\,III] emission-line luminosities and radio
luminosities of 206 FRII radio sources is shown in
Figs.~\ref{OIII-hist} and ~\ref{151-hist}. There are 10 radio galaxies
and two WQs with power $<2\times10^{25}$ W Hz$^{-1}$ sr$^{-1}$,
typical for the transition region for FRI and FRII sources. Quasars
appear to be more luminous than HEGs and WQs in both [O\,III]
emission-lines and radio. The Kolmogorov-Smirnov test
(Table~\ref{KS-test}) shows that the luminosity distributions
($L_{[\rm O\,III]}$ and $L_{151}$) of quasars are different from the
distributions of HEGs and WQs at high confidence level (mainly because
of the lack of weak quasars in the sample), whilst the $L_{[\rm
O\,III]}$ and $L_{151}$ distributions of HEGs and WQs appear to be
drawn from the same population. The later result is supported by the
linear-size statistics of FRII sources: the mean projected linear
size of weak quasars, ($330\pm50$) kpc, is comparable with the mean
linear size of narrow-line HEGs, ($320 \pm 30$) kpc, and is twice as
large as the mean size of quasars ($162 \pm 19$) kpc. This
relation holds over the entire [O III] emission-line luminosity range
(Fig.~\ref{size_LOIII}). The values of mean linear sizes of weak
quasars are larger than those of quasars in three equal subsamples
divided by [O III] emission-line luminosity. The same result is
obtained when the sample is binned by 151 MHz radio luminosity. This
supports the idea that HEGs and weak quasars are the same objects,
where the later ones appear to have broad emission lines as a result
of scattered light from the nucleus. Therefore in this analysis,
the weak quasars are grouped with the HEGs rather than quasars. On
the other hand, some of the WQs appear to be transitional objects
(between quasars and HEGs) which are oriented near to the critical
angle where the broad-line region is reddened but not totally obscured
(Laing et al. 1994; Simpson et al. 1999) and hence they can be grouped
with quasars. Therefore it is important to see how an exclusion of 20
WQs influences the final results and this will be discussed later in
this section.

For the whole sample, I calculate the number of all radio sources
$N_{\rm G+Q}=206$ and quasars $N_{\rm Q}=52$, and their mean linear
sizes $\overline{l}=(279 \pm 21)$ kpc and $\overline{l}_{\rm Q}=(162
\pm 19)$ kpc respectively. Then the critical viewing angles,
${\theta}_{\rm cs}=42^{\circ}\pm7^{\circ}$ and ${\theta}_{\rm
cf}=42^{\circ}\pm12^{\circ}$, are estimated using equations (7) and
(1). Two independent equations give similar results with the error
being smaller for the mean linear size statistics (see Appendix for
explanation). The true value of ${\overline{\theta}}_{\rm cs}$ lies
between $\sim 20^{\circ}$ and $\sim 60^{\circ}$ at $3\sigma$ (99.7 \%)
confidence level. There is a spread in the values of the critical
viewing angle, which indicates that the simple unification scheme with
constant $\theta_{\rm c}$ is not likely - there is some distribution
of $\theta_{\rm c}$. It would be interesting to see if the spread in
$\theta_{\rm c}$ correlates with any other physical property of radio
sources.

   \begin{figure} \centering
   \includegraphics[angle=-90,width=8.2cm]{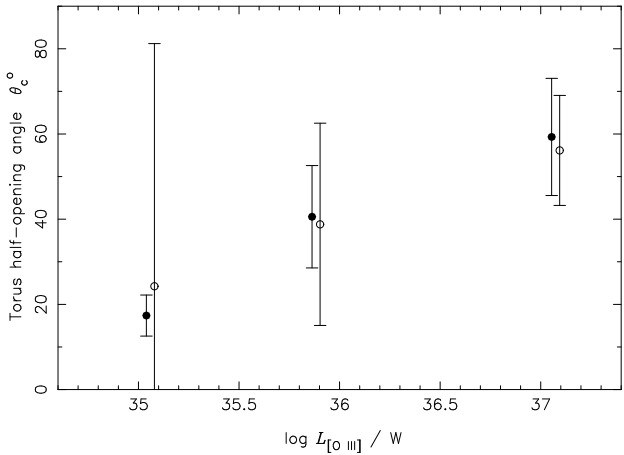}
   \caption{The torus half opening angle calculated in three equal
   subsamples binned by [O III] emission-line luminosity (see
   Table~\ref{F-test}). Critical angles ${\theta}_{\rm cs}$ estimated
   by the mean linear sizes of radio galaxies and quasars (eq. 7) are
   marked by filled circles, and open circles denote the critical
   angles ${\theta}_{\rm cf}$ calculated by the fraction of quasars
   (eq. 1). A $Fisher$ test rejects the null hypothesis that
   ${\theta}_{\rm cs}$ are equal at 97.4 \% confidence level. The
   filled circles are offset horizontally for illustrative
   purposes. $1\sigma$ error bars are presented.}  \label{OIII-theta}
   \end{figure}
%

To examine the $\overline{\theta}_{\rm c}-L_{[\rm O\,III]}$
dependence, I divided the sample into two equal subsamples (118 and
119 sources) by $\log (L_{[\rm O\,III]} / \rm{W})=35.83$ and
calculated $\overline{\theta}_{\rm cs}$ and $\overline{\theta}_{\rm
cf}$ for each subsample (Table~\ref{F-test}). Both,
$\overline{\theta}_{\rm cs}$ and $\overline{\theta}_{\rm cf}$,
increase by a factor of two times from low to high $L_{[\rm O\,III]}$.
The $Fisher$ test is used to test the null hypothesis that the mean
critical angles are equal at low and high $L_{[\rm O\,III]}$. For
${\theta}_{\rm cs}$ the null hypothesis is rejected at high confidence
level (99 \%), though it can not be rejected for ${\theta}_{\rm cf}$
(49 \%) mainly because of the large errors involved (see Appendix). I
also studied the $\overline{\theta}_{\rm c}-L_{[\rm {O\,III]}}$ plane
for three equal subsamples (Table~\ref{F-test};
Fig.~\ref{OIII-theta}). The mean half opening angle increases from
$\sim 20^{\circ}$ to $\sim 40^{\circ}$ and to $\sim 60^{\circ}$ in
three subsamples binned by [O III] emission-line luminosity, and the
difference between the angles is still very significant (97 \%). The
difference becomes less significant ($<85$ \%) when the sample is
divided on four or more subsamples. The division into subsamples
smaller than $\sim 50$ objects does not allow a statistically
significant correlation between $\overline{\theta}_{\rm c}$ and
$L_{[\rm O\,III]}$ to be investigated. Also, the small samples may
introduce fluctuations of $\overline{\theta}_{\rm c}$
due to statistical fluctuations in the orientation of radio
sources (Saikia \& Kulkarni 1994).

   \begin{table}
     \caption[]{The significance ($Fisher$ test) of the null
       hypothesis that the ${\theta}_{\rm cs}$ are equal in two equal
       subsamples divided by $L_{[\rm {O\,III}]}$, $L_{\rm {151}}$ and
       redshift $z$, respectively. Probabilities are calculated for:
       (1) the sample which includes 20 WQs grouped with HEGs rather
       than quasars, (2) the sample consisting only of HEGs and
       quasars, and (3) the sample which excludes 12 transitional
       sources with $\log(L_{151}$ / W Hz$^{-1}$ sr$^{-1}) < 25.3$.}
     \label{F-test1}
     $$ 
     \begin{tabular}{lccc}
       \hline
       \noalign{\smallskip}
       Relation & {$Fisher$\,test} (\%) & {$Fisher$\,test} (\%) & $Fisher$\,test (\%) \\
       plane    &   (1)           &    (2)          &   (3)  \\
       \noalign{\smallskip}
       \hline
       \noalign{\smallskip}
       $\overline{\theta}_{\rm cs}-L_{[\rm {O III}]}$ & 99.4 & 99.3 & 98.6 \\
       $\overline{\theta}_{\rm cs}-L_{\rm {151}}$     & 63.6 & 71.7 & 60.2 \\
       $\overline{\theta}_{\rm cs}-z$                 & 4.8  & 32.5 & 38.7 \\
       \noalign{\smallskip}
       \hline
     \end{tabular}
     $$ 
   \end{table}

   \begin{figure}
   \centering
   \includegraphics[angle=-90,width=8.2cm]{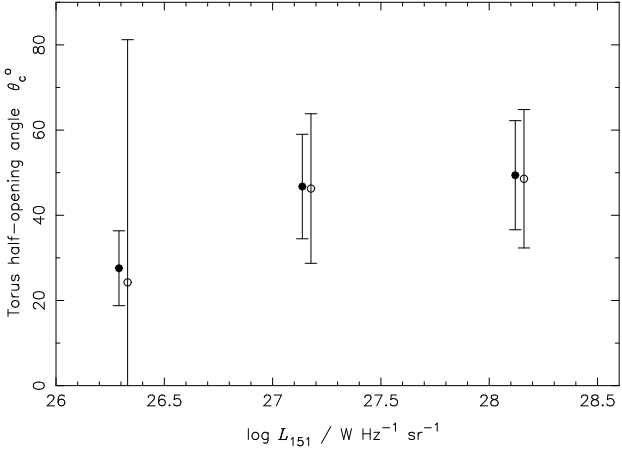}
      \caption{The half opening angle of the torus calculated in three
      equal subsamples binned by radio luminosity. Designations are
      the same as in Fig.~\ref{OIII-theta}. The null hypothesis of
      equal critical angles ${\theta}_{\rm cs}$ (filled circles) can
      not be rejected (66\%).}
         \label{L151-theta}
   \end{figure}
%

I repeated the analysis for the $\overline{\theta}_{\rm c}-L_{151}$
and $\overline{\theta}_{\rm c}-z$ relations. Both
$\overline{\theta}_{\rm cs}$ and $\overline{\theta}_{\rm cf}$
estimated for two equal subsamples binned by $L_{151}$ increase from
low- to high-luminosities. This result was expected because the
relation $L_{[\rm O\,III]} \propto L^{1.04}_{151}$ holds for the
combined 3CRR, 6CE and 7CRS samples (Grimes et al. 2004). A $Fisher$
test shows that the increase of $\overline{\theta}_{\rm cs}$ with
$L_{151}$ is not statistically significant (Table~\ref{F-test1};
Fig.~\ref{L151-theta}). No significant evolution of
$\overline{\theta}_{\rm cs}$ or $\overline{\theta}_{\rm cf}$ is found
with cosmic epoch $z$.

The results of this section stand even after excluding the WQs from
the sample (Table~\ref{F-test1}). This indicates that the positive
correlation in the $\overline{\theta}_{\rm cs}-L_{[\rm O\,III]}$ plane
is mainly due to HEGs and quasars. The significance of the
correlation is still high after exclusion of transitional sources from
the sample of 206 radio sources.

Let us consider the selection effects which might cause the positive
correlation $\overline{\theta}_{\rm c}-L_{[\rm O\,III]}$ seen in
Fig.~\ref{OIII-theta}. Although there are problems with identifying
the broad-line radio galaxies as low-luminosity quasars, Hardcastle et
al. (1998) suggest that some of the low-luminosity broad-line radio
galaxies are true quasars which are not sufficiently bright to be
classified as such in the optical. If so, then reassigning them to the
HEGs rather than to the quasars will decrease the fraction of quasars,
resulting in small opening angles and the consequence of this is that
$\overline{\theta}_{\rm cf}=24^{\circ}$ is underestimated in the
$L_{[\rm O\,III]}$ low-luminosity bin (Table~\ref{F-test}). In a
$L_{[\rm O\,III]}$ high-luminosity bin the fraction of quasars may
increase if the [O III] emitting region is partially obscured in radio
galaxies causing a factor of 5 to 10 depression of the [O III]
emission (Hes et al. 1996) which may result in the overestimation of
$\overline{\theta}_{\rm cf}$. The selection effects at low- and
high-luminosities may reproduce a positive correlation between
$\overline{\theta}_{\rm cf}$ and $L_{[\rm O III]}$. To understand the
bias introduced in the fraction of quasars one may consider the
independent measurements of the opening angle $\overline{\theta}_{\rm
cs}$ (Table~\ref{F-test}). The fact that $\overline{\theta}_{\rm cs}$
is independent of the fraction of quasars (eq. 7),
$\overline{\theta}_{\rm cs}$ positively correlates with $L_{[\rm
O\,III]}$ (Fig.~\ref{OIII-theta}) and $\overline{\theta}_{\rm cs}
\approx \overline{\theta}_{\rm cf}$ across $L_{[\rm O\,III]}$ range
indicates that the $\overline{\theta}_{\rm cf}-L_{[\rm O\,III]}$
correlation is real and if there is a bias in the fraction of quasars
it is not significant. The $\chi^2$ test is used to estimate the
significance of the null hypothesis that two independent measurements
of ${\overline{\theta}}_{\rm cs}$ and ${\overline{\theta}}_{\rm cf}$
are equal across the $L_{[\rm O\,III]}$ and $L_{151}$ ranges. For two
and three sets of binned data the values of ${\overline{\theta}}_{\rm
cs}$ and ${\overline{\theta}}_{\rm cf}$ are not significantly
different ($P<80$ \%). Figs.~\ref{OIII-theta} and ~\ref{L151-theta}
show that ${\overline{\theta}}_{\rm cs} \simeq
{\overline{\theta}}_{\rm cf}$ in different bins of $L_{[\rm O\,III]}$
and $L_{151}$. The equality of two independent estimates confirms the
validity of the orientation-based unification scheme for radio
galaxies and quasars and it indicates that the relative mean size of
quasars ($\overline{l}_{\rm Q}/\overline{l}$) and the relative number
of quasars ($N_{\rm Q}/N_{\rm G+Q}$) are indeed good measures of the
critical viewing angle in the low-frequency radio samples.

Another possible selection effect is related to the measured
projected sizes of FRII sources which may be overestimated as a result
of projection of the volume of radio lobes. It is clear that
independent of the plausible geometry of radio lobes the projection
effects may be important for sources having small inclinations of
radio axes to the line of sight. The fact that the exclusion of small
sizes ($< 20$ kpc, see section 3) does not change the final results
indicates that the projection effect is not likely to influence
strongly the estimates of $\theta_{\rm cs}$.

The assumption that the axis of a dusty torus is aligned with the
radio axis of FRII sources can be partially relaxed (e.g., Capetti \&
Celotti 1999) in real life allowing some distribution of angles
between the axes of the torus and the jet. In this scenario there
should be a difference between the true critical angle ($\theta_{\rm
ct}$) defined by the geometry of the torus and the critical angle
$\theta_{\rm cs}$ (eq. 7) estimated by means of projected sizes of
FRII sources. It is important to understand how this difference varies
depending on the value of true critical angle and whether it can lead
to a positive correlation found between $\theta_{\rm cs}$ and $L_{[\rm
O\,III]}$. To test this, I model random angles between the
axes of the torus and the jet by generating a pair of 5000 random
directions: for every random axis of the torus the random jet
direction is generated within the specified cone angle $2\delta_{\rm
max}$. For simplicity, a single linear size (300 kpc) for radio
galaxies and quasars is adopted. Then, given the true critical angle
($\theta_{\rm ct}$), I generate the projected linear sizes for both
quasars and all sources using the inclination angles of the jet. The
mean sizes of quasars and the whole sources are used to estimate the
critical angle $\theta_{\rm cs}$ (eq. 7). The deviation of estimated
critical angle $ \theta_{\rm cs}$ from the true $\theta_{\rm ct}$ one
is presented in Table~\ref{simul} for different $\delta_{\rm
max}$. $\delta_{\rm max}=0$ corresponds to the case where the axes are
aligned and here there is an excellent agreement between $\theta_{\rm
ct}$ and $\theta_{\rm cs}$, which provides a good test of the formalism
involved in equation (7). For larger values of $\delta_{\rm max}$ the
estimated critical angle appears to be larger than the true critical
angle ($\theta_{\rm cs} > \theta_{\rm ct}$) (Table~\ref{simul}). One
may conclude that the effect of misalignment of axes of the torus and
the jet leads to the overestimation of critical angles $\theta_{\rm
cs}$, which is higher for small $\theta_{\rm ct}$. If there is a
significant misalignment of axes then the critical angles in the
$\theta_{\rm cs} - L_{[\rm O\,III]}$ relation plane are overestimated
with the effect of making flatter the slope of the real $\theta_{\rm
ct} - L_{[\rm O\,III]}$ relation. This demonstrates that the
misalignment effect makes the real correlation weaker and thus can not
produce the positive correlation found between $\theta_{\rm cs}$ and
$L_{[\rm O\,III]}$.

\begin{table*}
\caption{The values of true critical angles $\theta_{\rm ct}$ and
their deviations ($\theta_{\rm cs}$) simulated for different half cone
angles $\delta_{\rm max}$ within which the angles between the axes of
the torus and the jet are distributed.}
\label{simul}
\begin{center} 
\begin{tabular} {l|cccc}
  
\hline $\theta_{\rm ct}^{\circ}$ & \multicolumn{4}{c}{$\theta_{\rm cs}^{\circ} \pm \sigma_{\theta_{\rm cs}}$} \\ 
\hline                   & $\delta_{\rm max}=0^{\circ}$ & $\delta_{\rm max}=10^{\circ}$ & $\delta_{\rm max}=20^{\circ}$ & $\delta_{\rm max}=30^{\circ}$ \\ 
                         & ($\theta_{\rm ct} = \theta_{\rm cs}$) & & & \\
\hline       20          & $19.8\pm0.4$ & $21.6\pm0.5$ & $26.5\pm0.7$ & $34.9\pm0.9$ \\ 
\hline       40          & $39.5\pm0.5$ & $40.2\pm0.5$ & $42.8\pm0.6$ & $47.0\pm0.7$ \\ 
\hline       60          & $59.3\pm0.6$ & $59.9\pm0.3$ & $61.2\pm0.7$ & $63.4\pm0.7$ \\ 
\hline
 
\end{tabular}
\end{center}
\end{table*}

\section{Discussion and conclusions}
The key result of the previous section is that the opening angle of
the ionizing cone of the torus becomes larger at high [O III]
emission-line luminosities. The half opening angle $\tan \theta_{\rm
c} = 2r/h$ depends on the inner radius, $r$, of the torus and its
height, $h$ (the thickness of the dusty torus). In extreme cases, the
positive correlation between $\theta_{\rm c}$ and $L_{[\rm O\,III]}$
can be interpreted as (i) a systematic increase of the radius of the
torus with $L_{[\rm O\,III]}$, or (ii) decrease of the height with
$L_{[\rm O III]}$. The first relation naturally follows from the
receding torus model (Lawrence 1991; Hill et al. 1996). In this model
(i) the ionizing radiation from the AGN evaporates the circumnuclear
dust forming the inner wall of the torus at a distance $r$ which
increases as $r \sim L^{0.5}_{[\rm O\,III]}$ (where $L_{[\rm O\,III]}$
is assumed to be a measure of the ionizing luminosity), and (ii) the
height of the torus $h$ is independent of $L_{[\rm O\,III]}$. There is
observational evidence in favour of an $r - L_{[\rm O\,III]}$
relation. Minezaki et. al. (2004) used multi-epoch observations of the
Seyfert 1 galaxy NGC 4151 to show that the time lag between the
optical and near-infrared light curves grows with optical luminosity
as $\log \triangle t \propto L^{0.5}$. They interpreted this as
thermal dust reverberation in an AGN to relate a $\triangle t$ to the
inner radius of the dusty torus ($\triangle t \propto r$). Willott et
al. (2002) found that 3C/6C quasars have higher submillimetre
luminosities by a factor $>2$ than radio galaxies. They argued that
this is in quantitative agreement with the receding torus model if
there is a positive correlation between optical and far-infrared
luminosities. Another piece of evidence comes from the studies of the
mid-infrared spectra of 3C radio sources (Freudling et al. 2003). They
showed that the bands of polycyclic aromatic hydrocarbons (PAHs) are
weak in broad-line radio galaxies and they are much stronger in
narrow-line radio galaxies of similar luminosities. They interpreted
this difference as a result of the heating radiation from the central
nucleus: the hard radiation destroys PAHs in the broad-line regions
and heats larger dust grains at intermediate distances. The spectra of
broad-line galaxies and quasars originate in the broad-line region
where only few PAHs survive. In the narrow-line radio galaxies this
region is obscured and the spectra are dominated by cooler dust at
larger distances where more PAHs can survive. Simpson \& Rawlings
(2000) showed that the near-infrared spectral slopes of 3CR quasars
are steeper for less luminous objects and that the fraction of
moderately obscured, red quasars decreases with increasing radio
luminosity. Both relations are in agreement with the receding torus
model.

As indicated by Simpson (2003) there are several lines of evidence
indicating that the height of the torus is not a strong function of
luminosity.  If so, the half opening angle of the torus $\tan
\theta_{\rm c} = 2r/h \propto L^{0.5}_{[\rm O\,III]}/h$ should
increase with [O III] emission line luminosity even if there is a
spread in $h$. This trend is in agreement with the positive
correlation found between the half opening angle of the torus and [O
III] emission-line luminosity (Fig.~\ref{OIII-theta}).  To fit the
$\theta_{\rm c} - L_{[\rm O\,III]}$ positive correlation, I used the
relation predicted by the receding torus model, $\tan\theta_{\rm c}
\propto L^{A}_{[\rm O\,III]}$, where $A$ is a free parameter. Using
the measurements of $\overline{\theta}_{\rm cs}$ and
$\overline{\theta}_{\rm cf}$, and $\overline{L}_{[\rm O\,III]}$ (see
Table~\ref{F-test}) I calculate the power-law indices $A_{\rm
s}=0.35 \pm 0.08$ and $A_{\rm f}=0.26\pm0.02$ corresponding to
$\overline{\theta}_{\rm cs}$ and $\overline{\theta}_{\rm cf}$
respectively. The value 0.35 is in agreement with the predicted value
of 0.5 ($r \propto L^{0.5}_{[\rm O\,III]}$) within $\sim 2\sigma$
significance level. It is likely that the use of the averaged critical
angles and small number of points result in a smoothing of the real
relation, which results in smaller values of $A_{\rm s}$ and $A_{\rm
f}$ compared with the predicted one, 0.5. Larger values of $A_{\rm
s}=0.44$ and $A_{\rm f}=0.27$ are found when the number of $L_{[\rm
O\,III]}$ bins are doubled. The range of $A\sim$(0.3 to 0.4) is a
plausible estimate for both $A_{\rm s}$ and $A_{\rm f}$. If these
values are correct then in order to fit the receding torus model one
should expect that the height of the torus also correlates positively
with the emission-line luminosity in the form $h \propto L^{B}_{[\rm
O\,III]}$ where $B\sim$(0.1 to 0.2). From the consideration of the
ratio of the near-infrared luminosity to the bolometric luminosity for
Palomar-Green quasars, Cao (2005) found significant correlation
between the torus thickness and the bolometric luminosity, $h \propto
L_{\rm bol}^{0.37 \pm 0.05}$, which is somewhat stronger than is
expected from my results. A larger sample of radio sources is needed
to confirm the increase of the torus thickness with photoionizing
radiation of the AGN.

The second relation, i.e. the negative correlation between $h$ and
$L_{[\rm O\,III]}$ which may be a cause for an increase of
${\overline{\theta}_{\rm c}}$ with $L_{[\rm O\,III]}$, seems to be
unrealistic since it requires the inner radius of the torus to be
independent of $L_{[\rm O\,III]}$. One may conclude that the growth of
the opening angle of the torus with increasing $L_{[\rm O\,III]}$ is a
direct evidence of the receding-torus-like structures around central
nuclei of FRII radio sources. This result apparently breaks the
degeneracy (multi-population or receding torus) in the interpretation
of data in favor of the receding torus model. \\

The main conclusions of this analysis are as follows:
\begin{enumerate}
  \item On the basis of an inverse problem approach, an equation for
  estimating the opening angle of the torus by means of the mean
  projected linear sizes of FRII radio galaxies and quasars has been
  derived. The small errors involved in estimation of the opening
  angle makes the linear size statistics a powerful tool for
  investigating the orientation-dependent structures in AGN.
  \item An application of this equation to the combined complete sample
  of FRII radio sources revealed the following:
    \begin{itemize}
      \item The mean half opening angle of the torus is $42^{\circ}
      \pm 7^{\circ}$ for the whole sample.
      \item Statistically significant positive correlation is found
	between the opening angle of the torus and [O III]
	emission-line luminosity. The mean half opening angle grows
	from $\sim 20^{\circ}$ to $\sim 60^{\circ}$ between $L_{[\rm
	O\,III]}\sim10^{35}$~W to $10^{37}$~W. This dependence
	favours the receding torus model and is fitted with the
	$\tan \theta_{\rm c} \propto L^{0.35}_{[\rm O\,III]}$
	relation.
      \item No significant changes of the opening angle are found
      either with radio luminosity or cosmic epoch.
      \item Two independent measurements of the half opening angle are
      correlated across [O III] emission-line and radio luminosity
      ranges. This supports the validity of the unified scheme of AGN
      and obscuring torus model. It also confirms that the mean
      relative size and relative number of quasars in low-frequency
      radio samples can serve as good estimators of the opening angle
      of an obscuring material.
    \end{itemize}
\end{enumerate}

\begin{acknowledgements}
  I am very grateful to Chris Simpson and Martin Hardcastle for many
  valuable comments, to Moshe Elitzur and Thomas Beckert for useful
  discussions and comments, to Steve Rawlings for kindly supplying the
  electronic version of the 6CE/7CR databases and for discussions, and
  to Alan Roy for useful comments and critical reading of a draft of
  the paper. I also thank the anonymous referee for careful
  reading of the manuscript and valuable suggestions, and I thank 
  the Alexander von Humboldt Foundation for the award of a Humboldt
  fellowship.
\end{acknowledgements}

\section*{APPENDIX: Equation for estimating the critical viewing angle}
Suppose, we have a population of double radio sources with radio axes
oriented randomly in space. The objects are classified as radio
galaxies if the viewing angle between the axis of the obscuring
`torus' and the line of the sight is larger than a critical angle
defined by a half opening angle of the torus, $\theta > \theta_{\rm
c}$, otherwise, if $\theta < \theta_{\rm c}$, they are viewed as radio
quasars. There is observational evidence that the axis of a radio jet
is aligned with the axis of the torus (Jaffe et al. 1993; Verdoes
Kleijn et al. 2001; Canalizo et al. 2003). Assume that radio axes of
FRII radio sources are aligned with the axes of obscuring tori. Then
the half opening angle of the torus is measured by the angle
($\theta_{\rm cs}$) between the radio axis and the critical angle
$\theta_{\rm c}$, i.e. $\theta_{\rm c}\approx\theta_{\rm cs}$. The
task is to derive an equation which allows $\theta_{\rm cs}$ to be
estimated given the distribution function of projected linear sizes of
radio galaxies and quasars, $f(l)$.

The projected linear size $l = l_{0} \sin \theta$, where
$l_{0}\in{[0,l_{\rm m}]}$ is the deprojected linear size ($l_{\rm m}$
being the largest linear size in the sample), and
$\theta\in{[0,\pi/2]}$ is the angle between the radio axis (or axis of
the torus) and the line of sight. Assume that torus axes are
distributed isotropically over the sky, $K(\theta)=\sin\theta$, then,
\begin{equation}
  f(l) = l \int\limits_{l}^{l_{\rm m}}
  \frac{F(l_{0})\,dl_{0}}{l_{0}\sqrt{l_{0}^{2} - l^{2}}},
\end{equation}
where $F(l_{0})$ is the distribution function of linear sizes. This
integral equation was first derived by Chandrasekhar \& Munch (1950)
for the problem of reconstruction of the distribution of stellar
rotation velocities.

The mean linear size of the whole sample is given by,
\begin{equation}
  \overline{l_{0}}=\frac{4}{\pi}\,\overline{l}.
\end{equation}
For quasars ($\theta < \theta_{\rm cs}$) the equation (2) can be
written as,
\begin{equation}
  f(l) = C_{\rm Q}\,l \int\limits_{l/\sin\theta_{\rm cs}}^{l_{\rm m}}
  \frac{F(l_{0})\,dl_{0}}{l_{0}\sqrt{l_{0}^{2} - l^{2}}}
  \,\,\,\,\,\,\,\,\,\, (l\le l_{\rm m}{\sin\theta_{\rm cs}}),
\end{equation}
where $C_{\rm Q} = 1/(1 - \cos\theta_{\rm cs})$. The equation for the
moments is,
\begin{equation}
  \nu_{\rm n} = C_{\rm Q}\, \nu_{\rm 0n} \int\limits_{0}^{l_{\rm
  m}\sin\theta_{\rm cs}}
  \frac{l^{n+1}\,dl}{l_{0}^{n+1}\sqrt{l_{0}^{2} - l^{2}}},
\end{equation}
and the first moment, which corresponds to the mean linear size of
quasars,
\begin{equation}
  (\overline{l_{0}})_{\rm Q} = 2\overline{l}_{\rm Q} \,\,\frac{1 -
  \cos\theta_{\rm cs}}{\theta_{\rm cs} - \sin\theta_{\rm
  cs}\cos\theta_{\rm cs}}.
\end{equation}
The radio galaxies and quasars have the same distribution of linear
sizes, hence $(\overline{l_0})_{\rm Q} = \overline{l_0}$. From (3) and
(6) one may recover the equation for estimating the critical viewing
angle,
\begin{equation}
  \frac{\pi}{2}\frac{\overline{l}_{\rm Q}}{\overline{l}} =
  \frac{\theta_{\rm cs}-\sin\theta_{\rm cs}\cos\theta_{\rm cs}}{1 -
  \cos\theta_{\rm cs}}.
\end{equation}
Given the standard errors of the mean projected linear sizes of
quasars $\sigma_{\overline{l}_{\rm Q}}$ and the whole population
$\sigma_{\overline{l}}$, one can estimate the standard error of the
critical angle,
\begin{equation}
  \sigma_{\theta_{\rm cs}}^{2} = {\left( \frac{\partial \theta_{\rm
  cs}}{\partial \overline{l}} \right)^2 \sigma^2_{\overline{l}}
  + \left( \frac{\partial \theta_{\rm cs}}{\partial \overline{l}_{\rm
  Q}} \right)^2 \sigma^2_{\overline{l}_{\rm Q}}},
\end{equation} 
where,
\begin{equation}
\frac {\partial \theta_{\rm cs}}{\partial \overline{l}} = \frac{2}{\pi}
\frac{(\theta_{\rm cs} - \sin\theta_{\rm cs}\cos\theta_{\rm cs})^2}{A
\overline{l}_{\rm Q}},
\end{equation}
\begin{equation}
\frac {\partial \theta_{\rm cs}}{\partial \overline{l}_{\rm Q}} = -
\frac{\pi}{2} \frac{(1 - \cos\theta_{\rm cs})^2}{A \overline{l}},
\end{equation}
\begin{equation}
A = \sin\theta_{\rm cs}\,(\theta_{\rm cs} - \sin\theta_{\rm
c}\cos\theta_{\rm cs}) - 2\sin^2\theta_{\rm cs}\,(1 - \cos\theta_{\rm
c}).
\end{equation}

If there is a spread in $\theta_{\rm cs}$ then eqs. (7-11) can be used
for estimating the mean half opening angle of the torus
$\overline{\theta}_{\rm cs}$ and its standard error
$\sigma_{\overline{\theta}_{\rm cs}}$.

{Let us investigate the errors associated with critical angles
($\theta_{\rm cs}$ and $\theta_{\rm cf}$) defined by the mean size and
fraction of quasars respectively (eqs. 7 and 1). For this, I adopt the
mean linear size of $N_{\rm{G+Q}}=206$ radio galaxies and quasars,
${\overline{l}}=(279\pm21)$ kpc (see section 4), and the mean size of
$N_{\rm{Q}}=52$ quasars, ${\overline{l}_{\rm Q}}=(162\pm19)$ kpc, from 
the combined sample of FRII radio sources (see section 3). I assume
that the ratios $\sigma_{\overline{l}}/{\overline{l}}=0.075$ and
$\sigma_{\overline{l}_{\rm Q}}/{\overline{l}}_{\rm Q}=0.099$ are
constant for any critical angle $\theta_{\rm cs}$. For any given
$\theta_{\rm cs}$, I calculate $\sigma_{{\theta}_{\rm cs}}$ from
eqs. (8-11) using the estimates of ${\overline{l}_{\rm Q}}$ (eq. 7)
and its standard error (provided that their inverse ratio is equal to
0.099). The $\sigma_{{\theta}_{\rm cs}}-\theta_{\rm cs}$ dependence is
shown in Fig.~\ref{error} (solid line). The relation between
$\sigma_{{\theta}_{\rm cf}}$ and $\theta_{\rm cf}$ (dashed line) is
drawn on the assumptions that (i)
$\sigma_{N_{\rm{G+Q}}}=\sqrt{N_{\rm{G+Q}}}\simeq14$ and
$\sigma_{N_{\rm {Q}}}=\sqrt{N_{\rm{Q}}}\simeq7$, and that (ii) the
ratios $\sigma_{N_{\rm{G+Q}}}/N_{\rm{G+Q}}=0.068$ and $\sigma_{N_{\rm
{Q}}}/N_{\rm{Q}}=0.13$ are unchanged over ${\theta}_{\rm cf}$. Given
the $\theta_{\rm cf}$, one may calculate,
\begin{equation}
  \sigma_{\theta_{\rm cf}}^{2} = {\left( \frac{\partial \theta_{\rm
  cf}}{\partial N_{\rm{G+Q}}} \right)^2 \sigma^2_{N_{\rm{G+Q}}} +
  \left( \frac{\partial \theta_{\rm cf}}{\partial N_{\rm{Q}}} \right)^2
  \sigma^2_{N_{\rm{Q}}}},
\end{equation} 
where,
\begin{equation}
\frac {\partial \theta_{\rm cf}}{\partial N_{\rm{G+Q}}} = - \frac{N_{\rm{Q}}}{N_{\rm{G+Q}}^2}
\left (1 - \left( 1-\frac{N_{\rm{Q}}}{N_{\rm{G+Q}}} \right)^2  \right)^{-1/2},
\end{equation}
\begin{equation}
\frac {\partial \theta_{\rm cf}}{\partial N_{\rm{Q}}} = \frac{1}{N_{\rm{G+Q}}}
\left (1 - \left( 1-\frac{N_{\rm{Q}}}{N_{\rm{G+Q}}} \right)^2  \right)^{-1/2}.
\end{equation}
One can see (Fig.~\ref{error}) that the errors, $\sigma_{{\theta}_{\rm
cf}}$ and $\sigma_{{\theta}_{\rm cs}}$, are comparable at angles
$>50^{\circ}$. At smaller angles, the $\sigma_{{\theta}_{\rm cf}}$
grows exponentially, reaching the value of $50^{\circ}$ at
${\theta}_{\rm c}=20^{\circ}$, whilst $\sigma_{{\theta}_{\rm cs}}$
decreases gradually to zero degrees. The important implication of
this analysis is that the errors associated with ${\theta}_{\rm cs}$
are small indicating that the linear size statistics is a powerful
tool for studying the correlations between the opening angle of the
torus and physical characteristics of AGN in the low-frequency radio
samples. This approach also can be used to estimate the opening angle
of structures (around the central engines of double radio sources)
showing the orientation-dependent physical/spectral characteristics.

   \begin{figure}
   \centering \includegraphics[angle=-90,width=8.2cm]{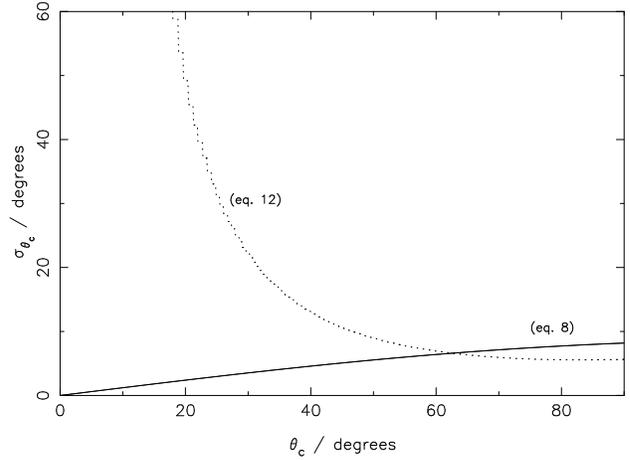}
      \caption{Functional relation between the half opening angle of
      the torus and its standard error is presented for equation (8)
      (solid line) and equation (12) (dotted line).}
         \label{error}
   \end{figure}
%

\end{document}